\documentclass{evn2024}
\usepackage{graphicx}
\usepackage{multirow}
\usepackage{natbib}
\usepackage{hyperref}

\DeclareRobustCommand{\VAN}[3]{#2}
\let\VANthebibliography\thebibliography
\def\thebibliography{\DeclareRobustCommand{\VAN}[3]{##3}\VANthebibliography}

\begin{document}
\title{Latest developments in wide-field VLBI}

\author{J.~F.~Radcliffe\inst{1,2}
		\and
		J.~P.~McKean\inst{2,3,4}
		\and 
		C.~Herbé-George\inst{2,3} 
		\and 
		L.~Coetzer\inst{2} 
		\and 
		T. Matsepane\inst{2}
		}

\institute{Jodrell Bank Centre for Astrophysics, University of Manchester, Oxford Road, Manchester M13 9PL, UK
		   \and
		   Department of Physics, University of Pretoria, Lynnwood Road, Hatfield, Pretoria 0083, South Africa
		   \and
		   Kapteyn Astronomical Institute, University of Groningen, Postbus 800, NL-9700 AV Groningen, The Netherlands
		   \and
		   South African Radio Astronomy Observatory (SARAO), P.O. Box 443, Krugersdorp 1740, South Africa
		   }

\abstract{Very Long Baseline Interferometry (VLBI) combines the signals of telescopes distributed across thousands of kilometres to provide some of the highest angular resolution images of astrophysical phenomena. Due to computational expense, typical VLBI observations are restricted to a single target and a small (few arcseconds) field-of-view per pointing. The technique of wide-field VLBI was born to enable the targeting of multiple sources and has been successful in providing new insights into Active Galactic Nuclei, the interstellar medium, supernovae, gravitational lenses and much more. However, this technique is still only employed in a few experiments, restricting the scientific potential of VLBI observations. In this conference proceeding, we outline new developments in wide-field VLBI, including an end-to-end correlation and calibration workflow, distributed correlation, and new calibration routines. These developments aim to enable wide-field VLBI to be a standard observing mode on all major VLBI arrays.}

\maketitle
%

\section{Introduction}

Very Long Baseline Interferometry (VLBI) combines the signals of radio telescopes distributed globally (and in space) to produce the highest angular resolution possible in all of astronomy. It has and continues to be used to make significant discoveries in many areas of astrophysics, including the direct imaging of a supermassive black hole's (SMBH) shadow \citep[e.g.,][]{2019ApJ...875L...1E,2022ApJ...930L..12E}, the localisation of Fast Radio Bursts \citep[e.g.,][]{2022Natur.602..585K}, mapping the ejecta of Tidal Disruption Events \citep[e.g.,][]{2018Sci...361..482M}, and mapping the structure of our own Galaxy \citep[e.g.,][]{2019ApJ...885..131R}.

However, most of these observations target a single object at the pointing centre, and an effect called time/bandwidth smearing reduces the effective field of view to \textit{sub-arcminute} diameter regions. This means that, for every VLBI observation, around 99 per cent of sources are not observed, and crucially, these sources cannot be recovered from the data products delivered to the user (at which point the original baseband data from the telescope is typically deleted).

As a result, wide-field VLBI was developed to observe multiple targets within a single observation. This was enabled through improvements in correlation \citep[e.g.,][]{2011PASP..123..275D} and calibration \citep[e.g.,][Coetzer et al. in prep.]{2013A&A...551A..97M,2016A&A...587A..85R} over the past twenty years. Wide-field VLBI has provided insights into a vast range of astrophysical fields, including Active Galactic Nuclei \citep[e.g.,][]{2001A&A...366L...5G,2013A&A...551A..97M,2017A&A...607A.132H,2018A&A...616A.128H,2018A&A...619A..48R,2023MNRAS.519.1732N,2024MNRAS.529.2428D,2024MNRAS.528.6141N}, galaxy evolution, star-formation \citep[e.g.,][]{2015MNRAS.452...32R,2019MNRAS.490.4024R}, supermassive black hole growth \citep[e.g.,][]{2017A&A...607A.132H}, the interstellar medium \citep[e.g.,][]{2013ApJ...768...12M}, gravitational lenses and cosmology \citep[e.g.,][]{2019MNRAS.483.2125S}, substellar and exoplanet radio emission \citep[e.g.,][]{2023Natur.619..272K,2023Sci...381.1120C} and young stellar objects \citep[e.g.,][]{2021ApJ...906...24D}. 

\begin{figure*}
    \centering
    \includegraphics[width=\linewidth]{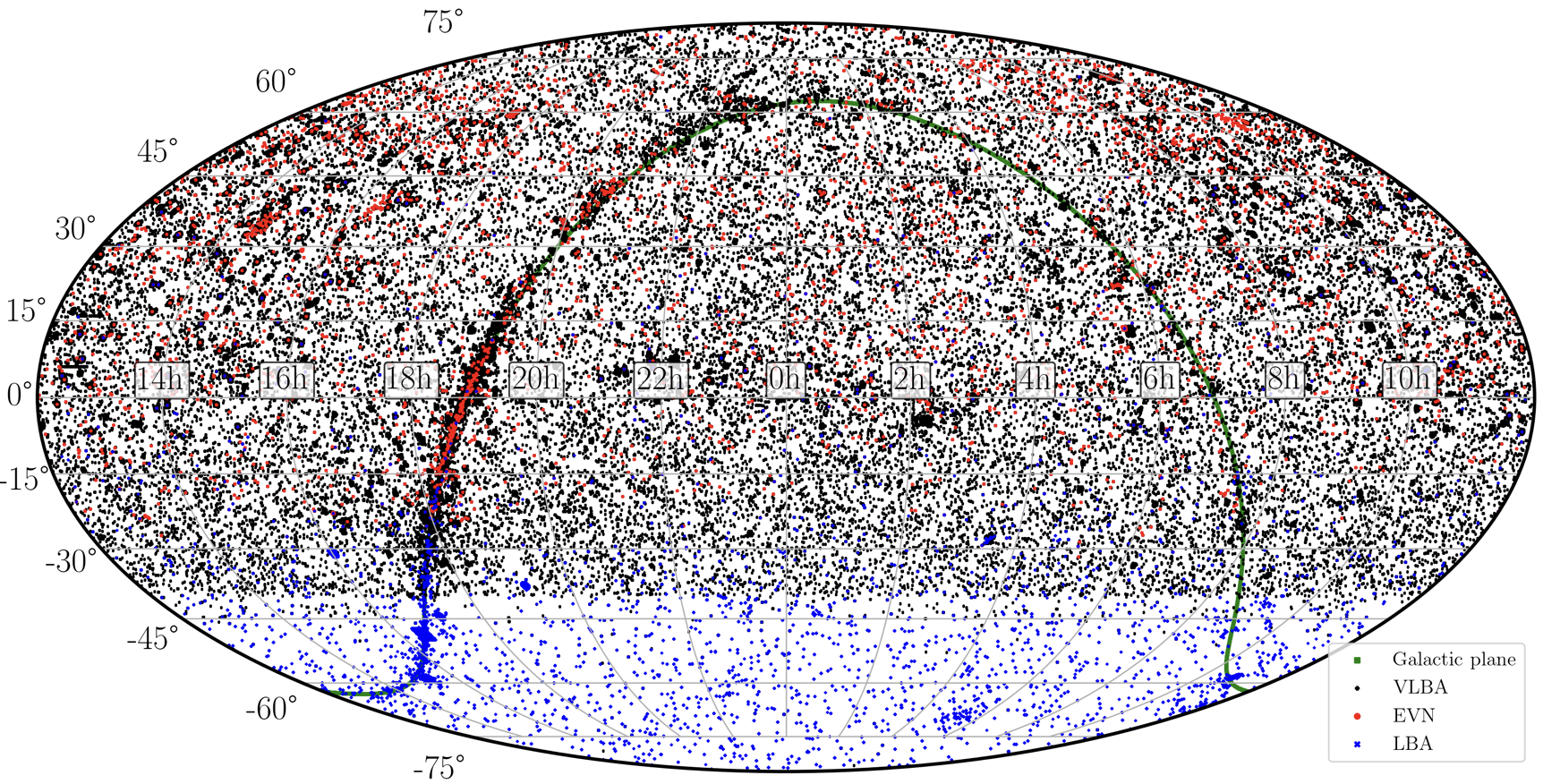}
    \caption{The pointings of all VLBI observations in the archives of the three major VLBI arrays: the European VLBI Network (EVN), Very Long Baseline Array (VLBA) and the Australian Long Baseline Array (LBA). The green solid line represents the galactic plane.}
    \label{f:all_sky_VLBI}
\end{figure*}

Despite numerous improvements in data processing techniques and computational capabilities, wide-field VLBI still needs to be utilised to its full potential, mainly due to its historically high computational complexity. Less than five per cent of observations currently use this technique, losing 95 per cent of sources and lots of possible science. This is highlighted in Figure~\ref{f:all_sky_VLBI}, which shows all VLBI observations of the three major arrays. Although there is dense, all-sky coverage, each observation covers a tiny fraction of the primary beam of the array, thus severely limiting the actual sky coverage. In this conference proceedings, we will outline the next generation of developments enabling us to realise wide-field VLBI as a standard observing mode on all VLBI arrays. 

\section{The road to all-sky VLBI}

As it stands, only a tiny fraction of VLBI experiments employ wide-field VLBI techniques. This means that many of the science cases outlined above will be limited in scope, and crucial science results will be missed. In light of this, there has been a push to develop commensal surveys that employ wide-field VLBI on top of standard PI-led experiments. While the scope is limited to pilot surveys (e.g., see Herbé-George et al., submitted), the ultimate goal is to conduct wide-field VLBI whenever a VLBI array observes the sky. Significant developments must enable wide-field VLBI to become the standard operating mode on all major VLBI arrays.

\begin{figure*}
	\centering
	\includegraphics[width=\linewidth]{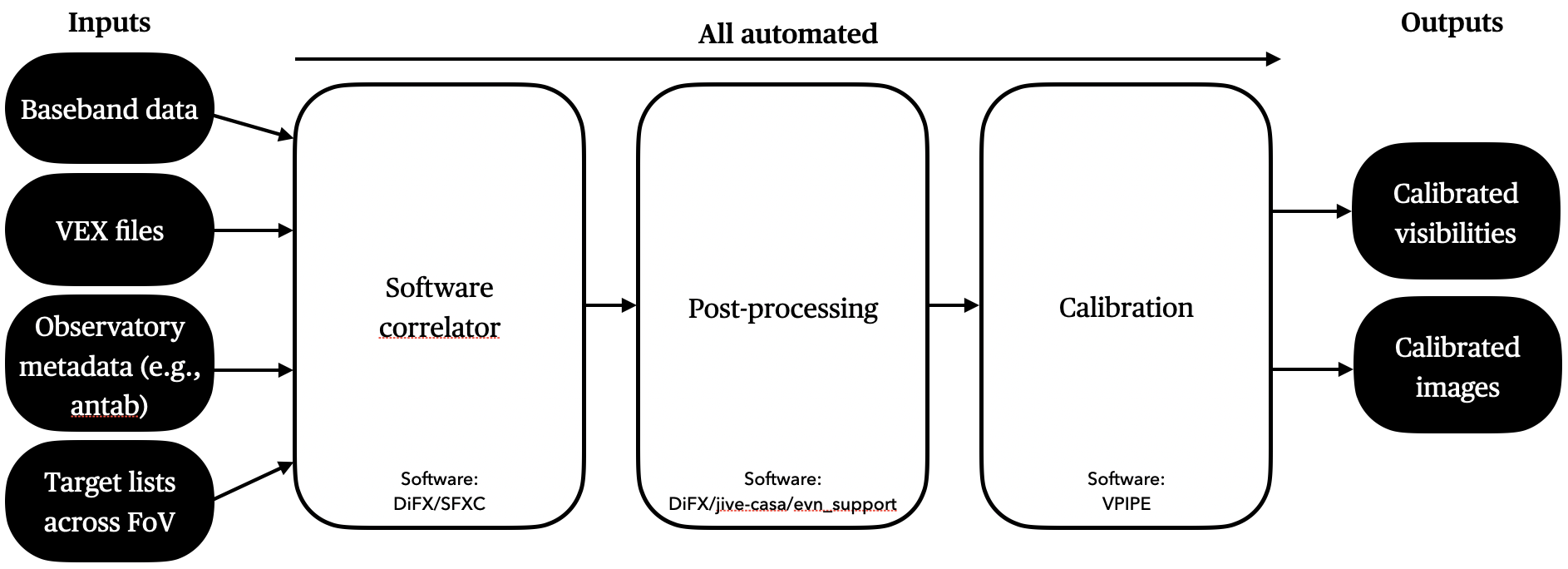}
	\caption{Block diagram of the end-to-end workflow developed in these commensal wide-field VLBI projects.}
	\label{f:commensal_workflow}
\end{figure*}

We have begun work on this through the development of the commensal wide-field VLBI workflow (see Figure~\ref{f:commensal_workflow}) to automate the transition from baseband VLBI data to correlated and calibrated visibilities and images across the whole field-of-view. The workflow comprises three main areas: i. software correlation, ii. post-processing, and iii. calibration. In the following sub-sections, we highlight some of the key advances in these areas (\S\ref{ss:s_corr} and \S\ref{ss:calib}), describe the control software and algorithms that implement this workflow (\S\ref{ss:imple}), and present some initial results in (\S\ref{ss:results}).

\subsection{Software correlation - distributed correlation}\label{ss:s_corr}

Correlation is a highly parallelisable process often distributed across multiple axes, such as time, frequency, polarisation, and baseline. Crucially, many of these axes are independent, i.e., a computation does not require information from other computations to complete. The current VLBI software correlators -- DiFX \citep{2011PASP..123..275D} and SFXC \citep{2015ExA....39..259K} -- do this only on a single cluster at a time, which limits the correlator's capabilities to the single cluster/server on which it is deployed. 

As a result, one of the most significant bottlenecks in conducting wide-field VLBI is the increased computational workload on the correlator cluster caused by targeting multiple targets within the field-of-view. From an operational standpoint, it is common sense to prioritise experiments requiring simple, standard and quick correlations, as many of these can be delivered to the PI in the same amount of time for which a single wide-field VLBI experiment can be processed. 

In our commensal program, we are exploring whether we can alleviate the restriction on the single cluster for correlation by providing a `distributed' correlation, i.e., leveraging multiple clusters simultaneously to access more compute resources. This has been investigated previously in the 2010--13 NEXPReS project\footnote{\href{https://www.nexpres.eu}{https://www.nexpres.eu}}, but limitations on data transport rates limited its usefulness at that time. However, the situation is different today, with 10 Gbps and above links available worldwide. This now means that the correlation can be distributed across individual remote clusters (even those that cannot pass information to each other) and run to completion. This should allow us to enable the order-of-magnitude increase in correlator output required to provide wide-field VLBI observations as a standard mode. We have begun work on a prototype that successfully shows distributed correlation across a server at the University of Pretoria and the Ilifu cloud computing cluster in Cape Town. We envisage an operational version to be available next year. 

\subsection{Calibration}\label{ss:calib}

Another area that has seen significant progress in recent years has been the calibration routines and pipelines for VLBI \citep[e.g., rPICARD;][]{Janssen2019}. As part of our end-to-end workflow, we have produced the VLBI pipeline \citep[VPIPE;][]{radcliffe2024}. This is a generic, flexible, and modular pipeline that is based on the Common Astronomy System Applications \citep[CASA;][]{casa2022} software and enabled by the VLBI developments within the software package \citep{2022PASP..134k4502V}. It is designed to calibrate GHz-frequency (tested on 1--22 GHz) phase-referenced continuum observations and currently supports EVN, VLBA, and LBA observations. It includes the ability to conduct self-calibration and includes wide-field specific routines, including processing multiple phase centres, multi-source self-calibration \citep{2016A&A...587A..85R} and primary beam corrections (Coetzer et al. in prep.). The software has been used in multiple publications and projects to date \citep[e.g.,][]{2023MNRAS.519.1732N,2024MNRAS.528.6141N,2024MNRAS.529.2428D}, and a complete publication is in preparation.

\subsection{Workflow implementation}\label{ss:imple}

These various developments are designed to be scalable and modular so that each part can be run individually or together. Furthermore, we required the software to be interoperable with heterogeneous computing systems, especially concerning distributed correlation. To enable this, we adopted the Apptainer software \citep{singularity}, which allows the required software to be portable and reproducible on different computing architectures. The execution of these workflows and software (outlined in Figure~\ref{f:commensal_workflow}) is implemented through some purpose-build control software\footnote{\href{https://github.com/jradcliffe5/control_widefield_vlbi}{https://github.com/jradcliffe5/control\_widefield\_vlbi}}. This software is still in the prototyping and development stage with the first full version expected to be ready in 2025.

\subsection{Initial results}\label{ss:results}

\begin{figure}
	\centering
	\includegraphics[width=\linewidth]{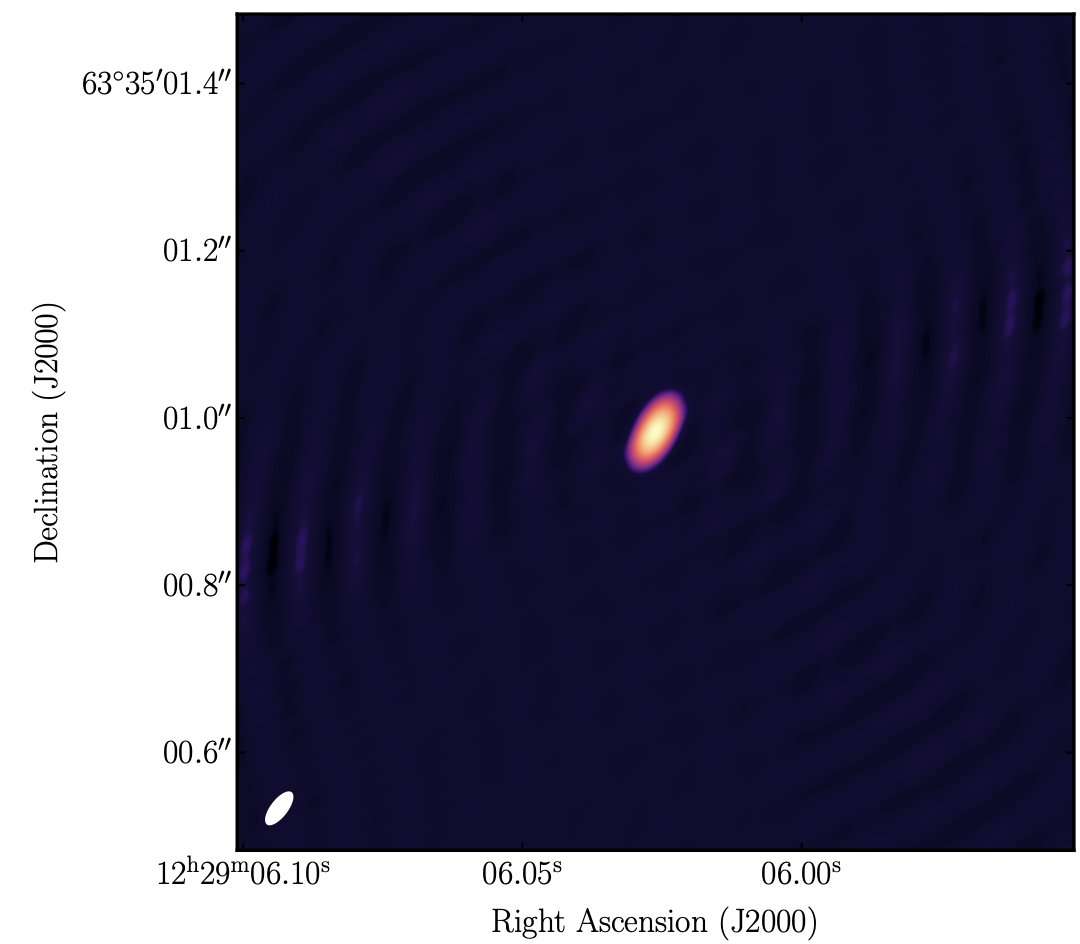}
	\caption{EVN image of source J1229+6335 at 1.6\,GHz, whose data was correlated and calibrated using the end-to-end correlation and calibration workflow shown in Figure~\ref{f:commensal_workflow}.}
	\label{f:first_smbh}
\end{figure}

To test these new developments, we applied the end-to-end workflow to a test EVN data set from the PRECISE\footnote{\href{http://www.ira.inaf.it/precise}{http://www.ira.inaf.it/precise}} collaboration (project code EK050E). This was chosen due to the small number of stations (7; Ef O8 Tr Ur Mc Nt Wb) and the standard frequency setup (18\,cm; 256\,MHz bandwidth). We correlated just 16 minutes of data on the phase reference source J1229+6335. For the correlation, we used SFXC, post-processed the data using the EVN support tools, and then calibrated these data using VPIPE. This was all done in a completely automated fashion with appropriate metadata, calibration strategies and correlation parameters derived using the VLBI EXperiment (VEX) file. The resulting image is shown in Figure~\ref{f:first_smbh}. The source is unresolved with a peak brightness of $0.4\,\mathrm{Jy\,beam^{-1}}$ and a 1$\sigma$ rms noise of $1.83\,\mathrm{mJy\,beam^{-1}}$, giving a signal-to-noise ratio of 218. 

This workflow is continuously developing and will be implemented for two separate programs. The first of these, the Synoptic Wide-field EVN-eMERLIN Public Survey (SWEEPS; PI: McKean), whose pilot survey is almost complete \citep[][Herbé-George et al., in prep.]{2025MNRAS.537L..49H}, will initially correlate one year's worth of L-band data from the European VLBI Network. This will allow us to refine this workflow and demonstrate the scientific value of these commensal surveys. The second program will correlate all L- and S-band data on the Australian Long Baseline Array (PIs: Collier \& Radcliffe), providing the Southern coverage of the VLBI sky, which is sorely missing (as shown in Figure~\ref{f:all_sky_VLBI}).

\section{Conclusions}

To conclude, wide-field VLBI provides unique insights into many phenomena that will be missed if the current VLBI observing mode remains the status quo. With the improvements in computational capabilities, the flexibility and scalability of current software correlators, coupled with the improvements such as distributed correlation and calibration techniques and pipelines, that were presented in this proceeding, we believe that it is the right time for VLBI arrays to consider implementing wide-field VLBI as their default observing mode. Furthermore, these tools can assist conventional PI observations through better calibration using the extra sources in the field of view and by lowering the barrier to using VLBI data by providing science-ready, calibrated data and images. The goal is ambitious but has the potential for colossal legacy value and the opportunity to explore a new, untouched parameter space full of discoveries.

\begin{acknowledgements}
JFR acknowledges support from the UK SKA Regional Centre (UKSRC). The UKSRC is a collaboration between the University of Cambridge, University of Edinburgh, Durham University, University of Hertfordshire, University of Manchester, University College London, and the UKRI Science and Technology Facilities Council (STFC) Scientific Computing at RAL. The UKSRC is supported by funding from UKRI STFC. JPM, TM, CH-G acknowledges support by the National Research Foundation of South Africa (Grant Number: 128943). The European VLBI Network (\href{www.evlbi.org}{www.evlbi.org}) is a joint facility of independent European, African, Asian, and North American radio astronomy institutes. Scientific results from data presented in this publication are derived from the following EVN project code(s): EK050 and EM164.
\textit{e}-MERLIN is a National Facility operated by the University of Manchester at Jodrell Bank Observatory on behalf of STFC. 
\end{acknowledgements}

\bibliographystyle{aa}
\bibliography{ref.bib}

\begin{thebibliography}{29}
\expandafter\ifx\csname natexlab\endcsname\relax\def\natexlab#1{#1}\fi

\bibitem[{{CASA Team} {et~al.}(2022){CASA Team}, {Bean}, {Bhatnagar}, {Castro}, {Donovan Meyer}, {Emonts}, {Garcia}, {Garwood}, {Golap}, {Gonzalez Villalba}, {Harris}, {Hayashi}, {Hoskins}, {Hsieh}, {Jagannathan}, {Kawasaki}, {Keimpema}, {Kettenis}, {Lopez}, {Marvil}, {Masters}, {McNichols}, {Mehringer}, {Miel}, {Moellenbrock}, {Montesino}, {Nakazato}, {Ott}, {Petry}, {Pokorny}, {Raba}, {Rau}, {Schiebel}, {Schweighart}, {Sekhar}, {Shimada}, {Small}, {Steeb}, {Sugimoto}, {Suoranta}, {Tsutsumi}, {van Bemmel}, {Verkouter}, {Wells}, {Xiong}, {Szomoru}, {Griffith}, {Glendenning}, \& {Kern}}]{casa2022}
{CASA Team}, {Bean}, B., {Bhatnagar}, S., {et~al.} 2022, \pasp, 134, 114501

\bibitem[{{Climent} {et~al.}(2023){Climent}, {Guirado}, {P{\'e}rez-Torres}, {Marcaide}, \& {Pe{\~n}a-Mo{\~n}ino}}]{2023Sci...381.1120C}
{Climent}, J.~B., {Guirado}, J.~C., {P{\'e}rez-Torres}, M., {Marcaide}, J.~M., \& {Pe{\~n}a-Mo{\~n}ino}, L. 2023, Science, 381, 1120

\bibitem[{{Deane} {et~al.}(2024){Deane}, {Radcliffe}, {Njeri}, {Akoto-Danso}, {Bernardi}, {Smirnov}, {Beswick}, {Garrett}, {Jarvis}, {Whittam}, {Bourke}, \& {Paragi}}]{2024MNRAS.529.2428D}
{Deane}, R.~P., {Radcliffe}, J.~F., {Njeri}, A., {et~al.} 2024, \mnras, 529, 2428

\bibitem[{{Deller} {et~al.}(2011){Deller}, {Brisken}, {Phillips}, {Morgan}, {Alef}, {Cappallo}, {Middelberg}, {Romney}, {Rottmann}, {Tingay}, \& {Wayth}}]{2011PASP..123..275D}
{Deller}, A.~T., {Brisken}, W.~F., {Phillips}, C.~J., {et~al.} 2011, \pasp, 123, 275

\bibitem[{{Dzib} {et~al.}(2021){Dzib}, {Forbrich}, {Reid}, \& {Menten}}]{2021ApJ...906...24D}
{Dzib}, S.~A., {Forbrich}, J., {Reid}, M.~J., \& {Menten}, K.~M. 2021, \apj, 906, 24

\bibitem[{{Event Horizon Telescope Collaboration} {et~al.}(2022){Event Horizon Telescope Collaboration}, {Akiyama}, {Alberdi}, {Alef}, {Algaba}, {Anantua}, {Asada}, {Azulay}, {Bach}, {Baczko}, {Ball}, {Balokovi{\'c}}, {Barrett}, {Baub{\"o}ck}, {Benson}, {Bintley}, {Blackburn}, {Blundell}, {Bouman}, {Bower}, {Boyce}, {Bremer}, {Brinkerink}, {Brissenden}, {Britzen}, {Broderick}, {Broguiere}, {Bronzwaer}, {Bustamante}, {Byun}, {Carlstrom}, {Ceccobello}, {Chael}, {Chan}, {Chatterjee}, {Chatterjee}, {Chen}, {Chen}, {Cheng}, {Cho}, {Christian}, {Conroy}, {Conway}, {Cordes}, {Crawford}, {Crew}, {Cruz-Osorio}, {Cui}, {Davelaar}, {De Laurentis}, {Deane}, {Dempsey}, {Desvignes}, {Dexter}, {Dhruv}, {Doeleman}, {Dougal}, {Dzib}, {Eatough}, {Emami}, {Falcke}, {Farah}, {Fish}, {Fomalont}, {Ford}, {Fraga-Encinas}, {Freeman}, {Friberg}, {Fromm}, {Fuentes}, {Galison}, {Gammie}, {Garc{\'\i}a}, {Gentaz}, {Georgiev}, {Goddi}, {Gold}, {G{\'o}mez-Ruiz}, {G{\'o}mez}, {Gu}, {Gurwell}, {Hada}, {Haggard}, {Haworth}, {Hecht}, {Hesper}, {Heumann}, {Ho}, {Ho}, {Honma}, {Huang}, {Huang}, {Hughes}, {Ikeda}, {Impellizzeri}, {Inoue}, {Issaoun}, {James}, {Jannuzi}, {Janssen}, {Jeter}, {Jiang}, {Jim{\'e}nez-Rosales}, {Johnson}, {Jorstad}, {Joshi}, {Jung}, {Karami}, {Karuppusamy}, {Kawashima}, {Keating}, {Kettenis}, {Kim}, {Kim}, {Kim}, {Kim}, {Kino}, {Koay}, {Kocherlakota}, {Kofuji}, {Koch}, {Koyama}, {Kramer}, {Kramer}, {Krichbaum}, {Kuo}, {La Bella}, {Lauer}, {Lee}, {Lee}, {Leung}, {Levis}, {Li}, {Lico}, {Lindahl}, {Lindqvist}, {Lisakov}, {Liu}, {Liu}, {Liuzzo}, {Lo}, {Lobanov}, {Loinard}, {Lonsdale}, {Lu}, {Mao}, {Marchili}, {Markoff}, {Marrone}, {Marscher}, {Mart{\'\i}-Vidal}, {Matsushita}, {Matthews}, {Medeiros}, {Menten}, {Michalik}, {Mizuno}, {Mizuno}, {Moran}, {Moriyama}, {Moscibrodzka}, {M{\"u}ller}, {Mus}, {Musoke}, {Myserlis}, {Nadolski}, {Nagai}, {Nagar}, {Nakamura}, {Narayan}, {Narayanan}, {Natarajan}, {Nathanail}, {Fuentes}, {Neilsen}, {Neri}, {Ni}, {Noutsos}, {Nowak}, {Oh}, {Okino}, {Olivares}, {Ortiz-Le{\'o}n}, {Oyama}, {{\"O}zel}, {Palumbo}, {Paraschos}, {Park}, {Parsons}, {Patel}, {Pen}, {Pesce}, {Pi{\'e}tu}, {Plambeck}, {PopStefanija}, {Porth}, {P{\"o}tzl}, {Prather}, {Preciado-L{\'o}pez}, {Psaltis}, {Pu}, {Ramakrishnan}, {Rao}, {Rawlings}, {Raymond}, {Rezzolla}, {Ricarte}, {Ripperda}, {Roelofs}, {Rogers}, {Ros}, {Romero-Ca{\~n}izales}, {Roshanineshat}, {Rottmann}, {Roy}, {Ruiz}, {Ruszczyk}, {Rygl}, {S{\'a}nchez}, {S{\'a}nchez-Arg{\"u}elles}, {S{\'a}nchez-Portal}, {Sasada}, {Satapathy}, {Savolainen}, {Schloerb}, {Schonfeld}, {Schuster}, {Shao}, {Shen}, {Small}, {Sohn}, {SooHoo}, {Souccar}, {Sun}, {Tazaki}, {Tetarenko}, {Tiede}, {Tilanus}, {Titus}, {Torne}, {Traianou}, {Trent}, {Trippe}, {Turk}, {van Bemmel}, {van Langevelde}, {van Rossum}, {Vos}, {Wagner}, {Ward-Thompson}, {Wardle}, {Weintroub}, {Wex}, {Wharton}, {Wielgus}, {Wiik}, {Witzel}, {Wondrak}, {Wong}, {Wu}, {Yamaguchi}, {Yoon}, {Young}, {Young}, {Younsi}, {Yuan}, {Yuan}, {Zensus}, {Zhang}, {Zhao}, {Zhao}, {Agurto}, {Allardi}, {Amestica}, {Araneda}, {Arriagada}, {Berghuis}, {Bertarini}, {Berthold}, {Blanchard}, {Brown}, {C{\'a}rdenas}, {Cantzler}, {Caro}, {Castillo-Dom{\'\i}nguez}, {Chan}, {Chang}, {Chang}, {Chang}, {Chang}, {Chen}, {Chilson}, {Chuter}, {Ciechanowicz}, {Colin-Beltran}, {Coulson}, {Crowley}, {Degenaar}, {Dornbusch}, {Dur{\'a}n}, {Everett}, {Faber}, {Forster}, {Fuchs}, {Gale}, {Geertsema}, {Gonz{\'a}lez}, {Graham}, {Gueth}, {Halverson}, {Han}, {Han}, {Hasegawa}, {Hern{\'a}ndez-Rebollar}, {Herrera}, {Herrero-Illana}, {Heyminck}, {Hirota}, {Hoge}, {Hostler Schimpf}, {Howie}, {Huang}, {Jiang}, {Jinchi}, {John}, {Kimura}, {Klein}, {Kubo}, {Kuroda}, {Kwon}, {Lacasse}, {Laing}, {Leitch}, {Li}, {Liu}, {Liu}, {Lin}, {Lu}, {Mac-Auliffe}, {Martin-Cocher}, {Matulonis}, {Maute}, {Messias}, {Meyer-Zhao}, {Monta{\~n}a}, {Montenegro-Montes}, {Montgomerie}, {Moreno Nolasco}, {Muders}, {Nishioka}, {Norton}, {Nystrom}, {Ogawa}, {Olivares}, {Oshiro}, {P{\'e}rez-Beaupuits}, {Parra}, {Phillips}, {Poirier}, {Pradel}, {Qiu}, {Raffin}, {Rahlin}, {Ram{\'\i}rez}, {Ressler}, {Reynolds}, {Rodr{\'\i}guez-Montoya}, {Saez-Madain}, {Santana}, {Shaw}, {Shirkey}, {Silva}, {Snow}, {Sousa}, {Sridharan}, {Stahm}, {Stark}, {Test}, {Torstensson}, {Venegas}, {Walther}, {Wei}, {White}, {Wieching}, {Wijnands}, {Wouterloot}, {Yu}, {Yu (于威)}, \& {Zeballos}}]{2022ApJ...930L..12E}
{Event Horizon Telescope Collaboration}, {Akiyama}, K., {Alberdi}, A., {et~al.} 2022, \apjl, 930, L12

\bibitem[{{Event Horizon Telescope Collaboration} {et~al.}(2019){Event Horizon Telescope Collaboration}, {Akiyama}, {Alberdi}, {Alef}, {Asada}, {Azulay}, {Baczko}, {Ball}, {Balokovi{\'c}}, {Barrett}, {Bintley}, {Blackburn}, {Boland}, {Bouman}, {Bower}, {Bremer}, {Brinkerink}, {Brissenden}, {Britzen}, {Broderick}, {Broguiere}, {Bronzwaer}, {Byun}, {Carlstrom}, {Chael}, {Chan}, {Chatterjee}, {Chatterjee}, {Chen}, {Chen}, {Cho}, {Christian}, {Conway}, {Cordes}, {Crew}, {Cui}, {Davelaar}, {De Laurentis}, {Deane}, {Dempsey}, {Desvignes}, {Dexter}, {Doeleman}, {Eatough}, {Falcke}, {Fish}, {Fomalont}, {Fraga-Encinas}, {Freeman}, {Friberg}, {Fromm}, {G{\'o}mez}, {Galison}, {Gammie}, {Garc{\'\i}a}, {Gentaz}, {Georgiev}, {Goddi}, {Gold}, {Gu}, {Gurwell}, {Hada}, {Hecht}, {Hesper}, {Ho}, {Ho}, {Honma}, {Huang}, {Huang}, {Hughes}, {Ikeda}, {Inoue}, {Issaoun}, {James}, {Jannuzi}, {Janssen}, {Jeter}, {Jiang}, {Johnson}, {Jorstad}, {Jung}, {Karami}, {Karuppusamy}, {Kawashima}, {Keating}, {Kettenis}, {Kim}, {Kim}, {Kim}, {Kino}, {Koay}, {Koch}, {Koyama}, {Kramer}, {Kramer}, {Krichbaum}, {Kuo}, {Lauer}, {Lee}, {Li}, {Li}, {Lindqvist}, {Liu}, {Liuzzo}, {Lo}, {Lobanov}, {Loinard}, {Lonsdale}, {Lu}, {MacDonald}, {Mao}, {Markoff}, {Marrone}, {Marscher}, {Mart{\'\i}-Vidal}, {Matsushita}, {Matthews}, {Medeiros}, {Menten}, {Mizuno}, {Mizuno}, {Moran}, {Moriyama}, {Moscibrodzka}, {M{\"u}ller}, {Nagai}, {Nagar}, {Nakamura}, {Narayan}, {Narayanan}, {Natarajan}, {Neri}, {Ni}, {Noutsos}, {Okino}, {Olivares}, {Ortiz-Le{\'o}n}, {Oyama}, {{\"O}zel}, {Palumbo}, {Patel}, {Pen}, {Pesce}, {Pi{\'e}tu}, {Plambeck}, {PopStefanija}, {Porth}, {Prather}, {Preciado-L{\'o}pez}, {Psaltis}, {Pu}, {Ramakrishnan}, {Rao}, {Rawlings}, {Raymond}, {Rezzolla}, {Ripperda}, {Roelofs}, {Rogers}, {Ros}, {Rose}, {Roshanineshat}, {Rottmann}, {Roy}, {Ruszczyk}, {Ryan}, {Rygl}, {S{\'a}nchez}, {S{\'a}nchez-Arguelles}, {Sasada}, {Savolainen}, {Schloerb}, {Schuster}, {Shao}, {Shen}, {Small}, {Sohn}, {SooHoo}, {Tazaki}, {Tiede}, {Tilanus}, {Titus}, {Toma}, {Torne}, {Trent}, {Trippe}, {Tsuda}, {van Bemmel}, {van Langevelde}, {van Rossum}, {Wagner}, {Wardle}, {Weintroub}, {Wex}, {Wharton}, {Wielgus}, {Wong}, {Wu}, {Young}, {Young}, {Younsi}, {Yuan}, {Yuan}, {Zensus}, {Zhao}, {Zhao}, {Zhu}, {Algaba}, {Allardi}, {Amestica}, {Anczarski}, {Bach}, {Baganoff}, {Beaudoin}, {Benson}, {Berthold}, {Blanchard}, {Blundell}, {Bustamente}, {Cappallo}, {Castillo-Dom{\'\i}nguez}, {Chang}, {Chang}, {Chang}, {Chen}, {Chilson}, {Chuter}, {C{\'o}rdova Rosado}, {Coulson}, {Crawford}, {Crowley}, {David}, {Derome}, {Dexter}, {Dornbusch}, {Dudevoir}, {Dzib}, {Eckart}, {Eckert}, {Erickson}, {Everett}, {Faber}, {Farah}, {Fath}, {Folkers}, {Forbes}, {Freund}, {G{\'o}mez-Ruiz}, {Gale}, {Gao}, {Geertsema}, {Graham}, {Greer}, {Grosslein}, {Gueth}, {Haggard}, {Halverson}, {Han}, {Han}, {Hao}, {Hasegawa}, {Henning}, {Hern{\'a}ndez-G{\'o}mez}, {Herrero-Illana}, {Heyminck}, {Hirota}, {Hoge}, {Huang}, {Impellizzeri}, {Jiang}, {Kamble}, {Keisler}, {Kimura}, {Kono}, {Kubo}, {Kuroda}, {Lacasse}, {Laing}, {Leitch}, {Li}, {Lin}, {Liu}, {Liu}, {Lu}, {Marson}, {Martin-Cocher}, {Massingill}, {Matulonis}, {McColl}, {McWhirter}, {Messias}, {Meyer-Zhao}, {Michalik}, {Monta{\~n}a}, {Montgomerie}, {Mora-Klein}, {Muders}, {Nadolski}, {Navarro}, {Neilsen}, {Nguyen}, {Nishioka}, {Norton}, {Nowak}, {Nystrom}, {Ogawa}, {Oshiro}, {Oyama}, {Parsons}, {Paine}, {Pe{\~n}alver}, {Phillips}, {Poirier}, {Pradel}, {Primiani}, {Raffin}, {Rahlin}, {Reiland}, {Risacher}, {Ruiz}, {S{\'a}ez-Mada{\'\i}n}, {Sassella}, {Schellart}, {Shaw}, {Silva}, {Shiokawa}, {Smith}, {Snow}, {Souccar}, {Sousa}, {Sridharan}, {Srinivasan}, {Stahm}, {Stark}, {Story}, {Timmer}, {Vertatschitsch}, {Walther}, {Wei}, {Whitehorn}, {Whitney}, {Woody}, {Wouterloot}, {Wright}, {Yamaguchi}, {Yu}, {Zeballos}, {Zhang}, \& {Ziurys}}]{2019ApJ...875L...1E}
{Event Horizon Telescope Collaboration}, {Akiyama}, K., {Alberdi}, A., {et~al.} 2019, \apjl, 875, L1

\bibitem[{{Garrett} {et~al.}(2001){Garrett}, {Muxlow}, {Garrington}, {Alef}, {Alberdi}, {van Langevelde}, {Venturi}, {Polatidis}, {Kellermann}, {Baan}, {Kus}, {Wilkinson}, \& {Richards}}]{2001A&A...366L...5G}
{Garrett}, M.~A., {Muxlow}, T.~W.~B., {Garrington}, S.~T., {et~al.} 2001, \aap, 366, L5

\bibitem[{{Herb{\'e}-George} {et~al.}(2025){Herb{\'e}-George}, {McKean}, {Morganti}, \& {Radcliffe}}]{2025MNRAS.537L..49H}
{Herb{\'e}-George}, C., {McKean}, J.~P., {Morganti}, R., \& {Radcliffe}, J.~F. 2025, \mnras, 537, L49

\bibitem[{{Herrera Ruiz} {et~al.}(2017){Herrera Ruiz}, {Middelberg}, {Deller}, {Norris}, {Best}, {Brisken}, {Schinnerer}, {Smol{\v{c}}i{\'c}}, {Delvecchio}, {Momjian}, {Bomans}, {Scoville}, \& {Carilli}}]{2017A&A...607A.132H}
{Herrera Ruiz}, N., {Middelberg}, E., {Deller}, A., {et~al.} 2017, \aap, 607, A132

\bibitem[{{Herrera Ruiz} {et~al.}(2018){Herrera Ruiz}, {Middelberg}, {Deller}, {Smol{\v{c}}i{\'c}}, {Norris}, {Novak}, {Delvecchio}, {Best}, {Schinnerer}, {Momjian}, {Dettmar}, {Brisken}, {Koekemoer}, \& {Scoville}}]{2018A&A...616A.128H}
{Herrera Ruiz}, N., {Middelberg}, E., {Deller}, A., {et~al.} 2018, \aap, 616, A128

\bibitem[{{Janssen} {et~al.}(2019){Janssen}, {Goddi}, {van Bemmel}, {Kettenis}, {Small}, {Liuzzo}, {Rygl}, {Mart{\'\i}-Vidal}, {Blackburn}, {Wielgus}, \& {Falcke}}]{Janssen2019}
{Janssen}, M., {Goddi}, C., {van Bemmel}, I.~M., {et~al.} 2019, \aap, 626, A75

\bibitem[{{Kao} {et~al.}(2023){Kao}, {Mioduszewski}, {Villadsen}, \& {Shkolnik}}]{2023Natur.619..272K}
{Kao}, M.~M., {Mioduszewski}, A.~J., {Villadsen}, J., \& {Shkolnik}, E.~L. 2023, \nat, 619, 272

\bibitem[{{Keimpema} {et~al.}(2015){Keimpema}, {Kettenis}, {Pogrebenko}, {Campbell}, {Cim{\'o}}, {Duev}, {Eldering}, {Kruithof}, {van Langevelde}, {Marchal}, {Molera Calv{\'e}s}, {Ozdemir}, {Paragi}, {Pidopryhora}, {Szomoru}, \& {Yang}}]{2015ExA....39..259K}
{Keimpema}, A., {Kettenis}, M.~M., {Pogrebenko}, S.~V., {et~al.} 2015, Experimental Astronomy, 39, 259

\bibitem[{{Kirsten} {et~al.}(2022){Kirsten}, {Marcote}, {Nimmo}, {Hessels}, {Bhardwaj}, {Tendulkar}, {Keimpema}, {Yang}, {Snelders}, {Scholz}, {Pearlman}, {Law}, {Peters}, {Giroletti}, {Paragi}, {Bassa}, {Hewitt}, {Bach}, {Bezrukovs}, {Burgay}, {Buttaccio}, {Conway}, {Corongiu}, {Feiler}, {Forss{\'e}n}, {Gawro{\'n}ski}, {Karuppusamy}, {Kharinov}, {Lindqvist}, {Maccaferri}, {Melnikov}, {Ould-Boukattine}, {Possenti}, {Surcis}, {Wang}, {Yuan}, {Aggarwal}, {Anna-Thomas}, {Bower}, {Blaauw}, {Burke-Spolaor}, {Cassanelli}, {Clarke}, {Fonseca}, {Gaensler}, {Gopinath}, {Kaspi}, {Kassim}, {Lazio}, {Leung}, {Li}, {Lin}, {Masui}, {Mckinven}, {Michilli}, {Mikhailov}, {Ng}, {Orbidans}, {Pen}, {Petroff}, {Rahman}, {Ransom}, {Shin}, {Smith}, {Stairs}, \& {Vlemmings}}]{2022Natur.602..585K}
{Kirsten}, F., {Marcote}, B., {Nimmo}, K., {et~al.} 2022, \nat, 602, 585

\bibitem[{Kurtzer {et~al.}(2021)Kurtzer, cclerget, Bauer, Kaneshiro, Trudgian, \& Godlove}]{singularity}
Kurtzer, G.~M., cclerget, Bauer, M., {et~al.} 2021, hpcng/singularity: Singularity 3.7.3

\bibitem[{{Mattila} {et~al.}(2018){Mattila}, {P{\'e}rez-Torres}, {Efstathiou}, {Mimica}, {Fraser}, {Kankare}, {Alberdi}, {Aloy}, {Heikkil{\"a}}, {Jonker}, {Lundqvist}, {Mart{\'\i}-Vidal}, {Meikle}, {Romero-Ca{\~n}izales}, {Smartt}, {Tsygankov}, {Varenius}, {Alonso-Herrero}, {Bondi}, {Fransson}, {Herrero-Illana}, {Kangas}, {Kotak}, {Ram{\'\i}rez-Olivencia}, {V{\"a}is{\"a}nen}, {Beswick}, {Clements}, {Greimel}, {Harmanen}, {Kotilainen}, {Nandra}, {Reynolds}, {Ryder}, {Walton}, {Wiik}, \& {{\"O}stlin}}]{2018Sci...361..482M}
{Mattila}, S., {P{\'e}rez-Torres}, M., {Efstathiou}, A., {et~al.} 2018, Science, 361, 482

\bibitem[{{Middelberg} {et~al.}(2013){Middelberg}, {Deller}, {Norris}, {Fotopoulou}, {Salvato}, {Morgan}, {Brisken}, {Lutz}, \& {Rovilos}}]{2013A&A...551A..97M}
{Middelberg}, E., {Deller}, A.~T., {Norris}, R.~P., {et~al.} 2013, \aap, 551, A97

\bibitem[{{Morgan} {et~al.}(2013){Morgan}, {Argo}, {Trott}, {Macquart}, {Deller}, {Middelberg}, {Miller-Jones}, \& {Tingay}}]{2013ApJ...768...12M}
{Morgan}, J.~S., {Argo}, M.~K., {Trott}, C.~M., {et~al.} 2013, \apj, 768, 12

\bibitem[{{Njeri} {et~al.}(2023){Njeri}, {Beswick}, {Radcliffe}, {Thomson}, {Wrigley}, {Muxlow}, {Garrett}, {Deane}, {Moldon}, {Norris}, \& {Kothes}}]{2023MNRAS.519.1732N}
{Njeri}, A., {Beswick}, R.~J., {Radcliffe}, J.~F., {et~al.} 2023, \mnras, 519, 1732

\bibitem[{{Njeri} {et~al.}(2024){Njeri}, {Deane}, {Radcliffe}, {Beswick}, {Thomson}, {Muxlow}, {Garrett}, \& {Harrison}}]{2024MNRAS.528.6141N}
{Njeri}, A., {Deane}, R.~P., {Radcliffe}, J.~F., {et~al.} 2024, \mnras, 528, 6141

\bibitem[{Radcliffe(2024)}]{radcliffe2024}
Radcliffe, J. 2024, jradcliffe5/VLBI\_pipeline: v1.1

\bibitem[{{Radcliffe} {et~al.}(2019){Radcliffe}, {Beswick}, {Thomson}, {Garrett}, {Barthel}, \& {Muxlow}}]{2019MNRAS.490.4024R}
{Radcliffe}, J.~F., {Beswick}, R.~J., {Thomson}, A.~P., {et~al.} 2019, \mnras, 490, 4024

\bibitem[{{Radcliffe} {et~al.}(2016){Radcliffe}, {Garrett}, {Beswick}, {Muxlow}, {Barthel}, {Deller}, \& {Middelberg}}]{2016A&A...587A..85R}
{Radcliffe}, J.~F., {Garrett}, M.~A., {Beswick}, R.~J., {et~al.} 2016, \aap, 587, A85

\bibitem[{{Radcliffe} {et~al.}(2018){Radcliffe}, {Garrett}, {Muxlow}, {Beswick}, {Barthel}, {Deller}, {Keimpema}, {Campbell}, \& {Wrigley}}]{2018A&A...619A..48R}
{Radcliffe}, J.~F., {Garrett}, M.~A., {Muxlow}, T.~W.~B., {et~al.} 2018, \aap, 619, A48

\bibitem[{{Rampadarath} {et~al.}(2015){Rampadarath}, {Morgan}, {Soria}, {Tingay}, {Reynolds}, {Argo}, \& {Dumas}}]{2015MNRAS.452...32R}
{Rampadarath}, H., {Morgan}, J.~S., {Soria}, R., {et~al.} 2015, \mnras, 452, 32

\bibitem[{{Reid} {et~al.}(2019){Reid}, {Menten}, {Brunthaler}, {Zheng}, {Dame}, {Xu}, {Li}, {Sakai}, {Wu}, {Immer}, {Zhang}, {Sanna}, {Moscadelli}, {Rygl}, {Bartkiewicz}, {Hu}, {Quiroga-Nu{\~n}ez}, \& {van Langevelde}}]{2019ApJ...885..131R}
{Reid}, M.~J., {Menten}, K.~M., {Brunthaler}, A., {et~al.} 2019, \apj, 885, 131

\bibitem[{{Spingola} {et~al.}(2019){Spingola}, {McKean}, {Lee}, {Deller}, \& {Moldon}}]{2019MNRAS.483.2125S}
{Spingola}, C., {McKean}, J.~P., {Lee}, M., {Deller}, A., \& {Moldon}, J. 2019, \mnras, 483, 2125

\bibitem[{{van Bemmel} {et~al.}(2022){van Bemmel}, {Kettenis}, {Small}, {Janssen}, {Moellenbrock}, {Petry}, {Goddi}, {Linford}, {Rygl}, {Liuzzo}, {Marcote}, {Bayandina}, {Schweighart}, {Verkouter}, {Keimpema}, {Szomoru}, \& {van Langevelde}}]{2022PASP..134k4502V}
{van Bemmel}, I.~M., {Kettenis}, M., {Small}, D., {et~al.} 2022, \pasp, 134, 114502

\end{thebibliography}

\end{document}